\documentclass[preprint,12pt,authoryear]{elsarticle}

\usepackage{amsmath,amssymb}
\usepackage{graphicx}
\usepackage{booktabs}
\usepackage{array}
\usepackage{multirow}
\usepackage{siunitx}
\usepackage{xcolor}
\usepackage{url}
\usepackage{xurl}
\usepackage[hidelinks]{hyperref}

\journal{Applied Energy / eTransportation}
\hypersetup{pageanchor=false}
\newcommand{\beb}{battery-electric bus}

\newcommand{\method}{WeatherRobustBus}

\newcommand{\soc}{state of charge}

\begin{document}

\begin{graphicalabstract}
\centering
\fbox{%
\begin{minipage}[c][2.1in][c]{0.92\textwidth}
\centering
\textbf{Graphical abstract placeholder.}\\[0.6em]
Real cold-wave weather $\rightarrow$ real GTFS bus blocks $\rightarrow$ physics-anchored
energy and uncertainty $\rightarrow$ block-failure envelope $\rightarrow$ cost-robust
intervention frontier.
\end{minipage}}
\end{graphicalabstract}

\begin{highlights}
\item A weather-to-timetable failure mode hidden by seasonal energy margins is identified and quantified.
\item Real hourly cold waves are injected into real GTFS electric-bus duties at city scale.
\item Cabin-heating energy is validated against an independent EnergyPlus reference with confidence intervals.
\item A forecast-triggered robust policy cuts mean cold-wave block-failure probability from 0.76 to 0.11.
\item A de-confounded ablation reveals opportunity charging as the dominant winter robustness lever.
\end{highlights}

\begin{frontmatter}

\title{When the Timetable Breaks: Physics-Anchored Scientific Machine Learning for Cold-Wave-Robust Battery-Electric Bus Operations}

\author[inst1]{Yifan Wang\corref{cor1}}
\ead{yifan.wang18@mail.mcgill.ca}
\cortext[cor1]{Corresponding author.}

\affiliation[inst1]{organization={Department of Mechanical Engineering, McGill University},
            city={Montreal},
            state={QC},
            postcode={H3A 2T7},
            country={Canada}}

\begin{abstract}
Cold-climate transit agencies are electrifying their fleets at scale while still
operating fixed public timetables, and winter exposes a failure mode that
seasonal energy margins cannot see. During a cold wave, cabin heating drains the
battery faster than scheduled layovers can replenish it, so later trips in the
same vehicle block start under-charged and a single cold day cascades into
timetable infeasibility. We present \method{}, an open-data framework that turns
this latent risk into a measurable, actionable quantity by injecting real hourly
weather into real transit duties and propagating cold-weather energy uncertainty
into block-level failure probability. The framework couples a transparent
traction and cabin-thermal backbone with a bounded, monotone residual ensemble,
and validates its cabin-heating component against an independent EnergyPlus
bus-cabin simulation driven by the same Toronto weather record. Against this
first-principles reference the model attains the lowest all-year error
(\SI{0.213}{kWh} root-mean-square error over \num{8760} hours) and remains
reliable in the out-of-support cold tail ($T\leq\SI{-12}{\celsius}$), where every
pure-machine-learning baseline degrades by \numrange{1.5}{4}$\times$ and the best
competitor reaches only \SI{1.055}{kWh}. Embedded in a Monte~Carlo
block-feasibility simulator over \num{60} real Toronto TTC vehicle blocks, the
model reveals a sharp weather-induced failure envelope. A forecast-triggered
robust policy that combines opportunity charging, a fuel-fired cabin-heating
bridge and modest buffering reduces mean cold-wave failure probability from
\num{0.759} to \num{0.112} across eight cold-wave days, and a de-confounded
ablation shows that opportunity charging is the dominant lever while the heater
is a low-cost complement. \method{} provides a reproducible pathway from weather
data to winter-resilience decisions for electric-bus fleets.
\end{abstract}


\end{frontmatter}
\hypersetup{pageanchor=true}

\section{Introduction}

Public bus electrification has crossed the line from demonstration to dependence.
The Toronto Transit Commission already operates the largest \beb{} (BEB) fleet in
North America and is scaling toward several hundred vehicles that must serve the
same fixed public timetables, in the same winters, as the diesel fleet they
replace \citep{Li2016BEBReview,Perumal2022BusScheduling}. This transition quietly
changes the central engineering question. It is no longer whether cold weather
raises energy consumption, a fact that is well established for electric vehicles
and transit buses \citep{Gu2025ColdBEB,Gao2017BEBEnergyConsumption}. The question
that now governs service reliability is sharper and operational: can a real cold
wave make a scheduled vehicle block infeasible, and if so, which deployable
intervention restores the timetable at the least cost?

Cold waves are uniquely punishing for BEB operation because the passenger cabin
converts ambient temperature directly into electrical demand. As temperature
falls, heat-pump efficiency collapses while door cycling, infiltration and
comfort requirements stay locked to the schedule. A block that is comfortably
feasible under a seasonal-average derating factor can tip into infeasibility when
several long trips, short layovers and cold hours align. The consequence is not a
single energy-hungry trip; once the \soc{} (SoC) crosses the reserve threshold,
later trips in the same block inherit the deficit and the failure propagates
through the timetable. This is a structurally different risk from the smooth
seasonal range loss that current planning tools represent.

Existing BEB planning and scheduling research provides powerful machinery for
charger siting, block assignment and robust optimization under uncertain travel
time or energy consumption
\citep{Avishan2023Scheduling,Liu2018FastCharging,BenTal2009RobustOptimization,Perumal2022BusScheduling}.
Integrated transport--energy simulators have likewise advanced, including
microscopic platforms and GTFS-based depot-charging models that couple urban transportation with charging infrastructure
\citep{Qian2024V2Sim,Hendriks2024IntegratedModel}. Across this literature,
however, energy enters as a fixed coefficient, a scenario value, or an abstract
uncertainty set, and cabin thermodynamics under a \emph{historical cold-wave time
series} is never the stimulus that drives schedule failure. As a result, these
tools cannot say when a particular real timetable breaks, how much
weather-attributable deficit accumulates, or which lever changes the failure
probability the most. That gap is the entry point for this work.

The modelling obstacle is also statistical, and it is exactly where naive
data-driven approaches fail. The coldest operating hours are the hours that
decide robustness, yet they are rare and lie outside the support of ordinary
training data. A black-box model can be accurate in mild conditions and then
diverge precisely where heat-pump collapse and cabin load dominate. Scientific
machine learning offers a principled remedy by constraining learned functions
with physical structure
\citep{Raissi2019PINNs,Karniadakis2021PIML,Willard2022HybridSciML}. For winter
bus operations the effective form of this idea is not a generic network with a
physics penalty; it is a grey-box model that keeps a transparent traction and
cabin-energy backbone, learns only the residual that data can support, and emits
calibrated uncertainty that can be propagated into schedule-failure probability
\citep{Willard2022HybridSciML,Lakshminarayanan2017DeepEnsembles}.

This paper presents \method{}, a reproducible framework that closes the loop from
weather to winter-resilience decisions. It is driven by real Toronto TTC GTFS
duties \citep{TorontoOpenData2026,GTFS2026} and NASA POWER hourly weather
\citep{NASAPOWER2026}; it builds an independent EnergyPlus single-zone bus-cabin
reference \citep{Crawley2001EnergyPlus,EnergyPlus2025}, validates the cold-tail
heating model against it, and propagates energy uncertainty through a block-level
Monte~Carlo feasibility simulator. Its output is not an error metric but an
operational picture: a failure envelope, a failure probability for each cold-wave
day, deficit kilometers, fuel and carbon accounting, and an intervention
cost frontier.

Three contributions follow. First, we define and quantify a \emph{weather-induced
timetable-failure envelope}, a direct mapping from historical cold-wave weather
and real GTFS blocks to block-failure probability and range-deficit severity.
Second, we develop an EnergyPlus-validated, physics-anchored Sci-ML energy model
that is the most accurate model over the full reference year, is stable across
random seeds, and remains reliable in the out-of-support cold tail where pure
machine learning collapses. Third, we convert prediction into decision value
through a de-confounded ablation over opportunity charging, fuel-fired heating
and schedule buffering, and show that the full \method{} policy reduces mean
cold-wave failure probability by about \SI{85}{\percent} relative to
weather-blind seasonal operation. Together these results give transit agencies a
quantitative, auditable route from a weather forecast to a winter operating plan.

\section{Methodology}

\subsection{Problem definition and framework overview}

The unit of analysis is a vehicle block, an ordered sequence of scheduled trips
and layovers assigned to one bus. Let block $b$ contain legs $i=1,\ldots,n_b$.
Each leg has distance $d_i$, scheduled duration $\tau_i$, departure time $t_i$,
ambient weather $w(t_i)$ and an optional chargeable layover before the next leg.
The task is to estimate the energy-demand distribution of every leg, propagate it
through the block SoC recursion, and choose an intervention that minimizes
realized failure risk and cost on a cold-wave day.

\begin{figure}[t]
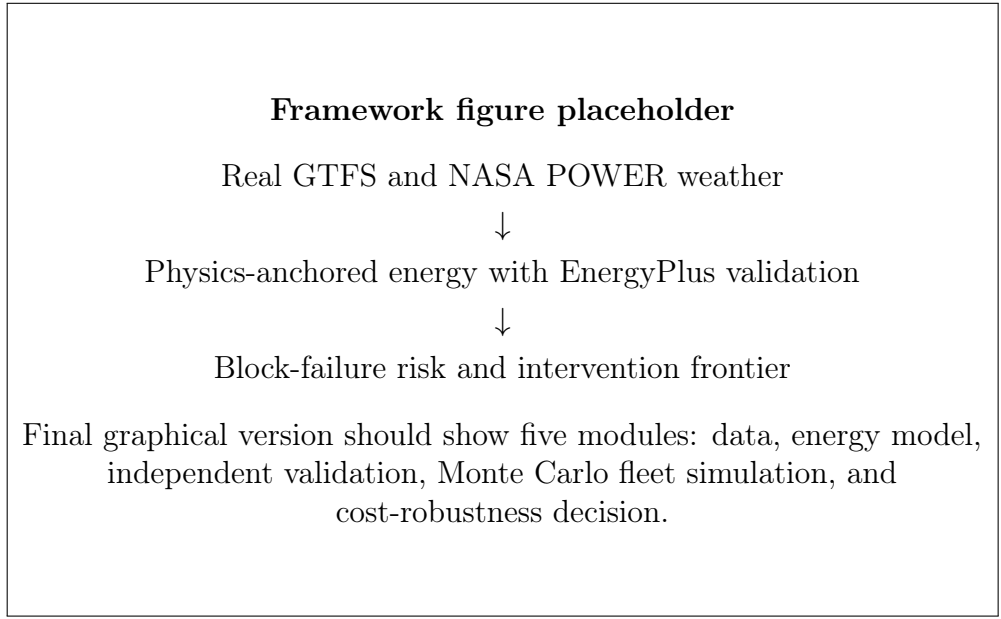

\centering
\fbox{%
\begin{minipage}[c][3.1in][c]{0.94\textwidth}
\centering
\textbf{Framework figure placeholder}\\[0.9em]
Real GTFS and NASA POWER weather\\[0.3em]
$\downarrow$\\[0.3em]
Physics-anchored energy with EnergyPlus validation\\[0.3em]
$\downarrow$\\[0.3em]
Block-failure risk and intervention frontier\\[0.9em]
Final graphical version should show five modules: data, energy model, independent validation,
Monte Carlo fleet simulation, and cost-robustness decision.
\end{minipage}}
\caption{Conceptual framework of \method{}. Historical hourly weather and real scheduled bus blocks enter a physics-anchored energy model, whose cabin-heating component is validated against an independent EnergyPlus simulation before its uncertainty is propagated into block-level failure risk and intervention decisions. The final artwork will replace this placeholder.}
\label{fig:framework}
\end{figure}

Figure~\ref{fig:framework} summarizes the framework. The input layer combines
Toronto TTC GTFS schedules with NASA POWER hourly weather. The energy layer maps
each leg and hour to a predictive energy distribution through a physics-anchored
residual ensemble. The validation layer compares the cabin-heating component to
an independent EnergyPlus reference. The operation layer samples leg energy,
advances SoC through each block and reports failure probability. The decision
layer evaluates three deployable interventions: opportunity charging, fuel-fired
auxiliary cabin heating and schedule buffering.

\subsection{Open data and simulation reference}

The transit input is the public TTC routes and schedules dataset, whose raw GTFS
feed contains \num{227} routes, \num{134647} trips and \num{6600} vehicle blocks.
For repeated Monte~Carlo evaluation, \num{60} real multi-trip blocks were sampled
with a fixed seed; these contain \num{1059} trips and \num{32429} stop-to-stop
legs. Leg distances were taken from GTFS shape-distance fields when available and
imputed from stop spacing only when missing. The sampled duties span
\SIrange{37}{455}{km}, with a median of approximately \SI{200}{km}, matching
realistic urban transit operation.

The weather input is NASA POWER hourly data at \SI{43.7001}{\degree}\,N,
\SI{79.4163}{\degree}\,W for 2019-01-01 to 2024-01-01, using 2~m air temperature, 2~m
wind speed and relative humidity. The processed record contains \num{43848}
hourly observations and \num{53} cold-wave windows, where a cold wave is a run of
consecutive days whose daily minimum temperature falls below the
10th percentile of all daily minima. The coldest day reached
\SI{-16.66}{\celsius}. The energy model was trained on $T\geq\SI{-2}{\celsius}$
and evaluated on the cold tail $T\leq\SI{-12}{\celsius}$, an explicit gap that
forces a genuine extrapolation test in the regime that governs failure.

The cabin-heating reference is an EnergyPlus~25.2 single-zone bus-cabin model: a
\SI{12.0}{m}$\times$\SI{2.55}{m}$\times$\SI{2.2}{m} zone with a lightweight
insulated envelope, long-wall glazing, \num{20} occupants, door-cycling
infiltration and an Ideal Loads Air System holding a \SI{19}{\celsius} heating
setpoint. The EPW weather file is generated from the same Toronto NASA POWER
record. EnergyPlus resolves the heating load through transient zone heat balance,
conduction transfer functions and psychrometric air balance, a formulation
deliberately distinct from the lumped $UA$ and coefficient-of-performance model
used by \method{}; agreement is therefore a genuine cross-model validation rather
than a self-consistency check. Table~\ref{tab:data} summarizes the data layers.

\begin{table}[t]
\centering
\caption{Data sources and processed scale used in the experiments.}
\label{tab:data}
\begin{tabular}{p{0.22\textwidth}p{0.30\textwidth}p{0.35\textwidth}}
\toprule
Layer & Source & Processed scale and role \\
\midrule
Transit timetable & Toronto TTC GTFS open data & 60 sampled blocks, 1059 trips and 32{,}429 legs from a full feed of 6600 blocks; basis for block-level feasibility. \\
Weather & NASA POWER hourly point API & 43{,}848 hourly records (2019--2024); 53 cold-wave windows; coldest daily minimum $-16.66^\circ$C. \\
Cabin-heating reference & EnergyPlus 25.2 with NASA-derived EPW & 8760 hourly heating-load records for the coldest weather year; independent validation target. \\
Vehicle and charging parameters & Documented simulation configuration & 350 kWh usable pack, 150 kW depot charging, 450 kW opportunity charging, SoC band 0.15--0.95. \\
\bottomrule
\end{tabular}
\end{table}

\subsection{Physics backbone for leg energy}

For each leg $i$, electrical energy is decomposed as
\begin{equation}
E_i^{\mathrm{leg}} =
E_i^{\mathrm{trac}} + E_i^{\mathrm{cab}} + E_i^{\mathrm{aux}} - E_i^{\mathrm{regen}},
\label{eq:energy_total}
\end{equation}
where $E_i^{\mathrm{trac}}$ is traction energy, $E_i^{\mathrm{cab}}$ is cabin
heating energy, $E_i^{\mathrm{aux}}$ is non-HVAC auxiliary demand and
$E_i^{\mathrm{regen}}$ is recovered braking energy.

The traction component follows a longitudinal force balance. For speed $v(t)$,
acceleration $a(t)$, road angle $\alpha(t)$, vehicle mass $m$, rolling
coefficient $C_{\mathrm{rr}}$, frontal area $A_f$, drag coefficient $C_d$ and air
density $\rho$, the wheel force is
\begin{equation}
F(t)=m a(t) + m g \sin \alpha(t) + C_{\mathrm{rr}} m g \cos \alpha(t)
      + \tfrac{1}{2}\rho A_f C_d v(t)^2 .
\label{eq:force}
\end{equation}
Positive wheel power discharges the battery through drivetrain efficiency
$\eta_{\mathrm{drv}}$, and negative wheel power is partially recovered through
regenerative efficiency $\eta_{\mathrm{reg}}$:
\begin{equation}
E_i^{\mathrm{trac}} - E_i^{\mathrm{regen}} =
\int_{t\in i} \left[\frac{F(t)v(t)}{\eta_{\mathrm{drv}}}\right]_+ \! dt
- \eta_{\mathrm{reg}}\!\int_{t\in i}[-F(t)v(t)]_+ \, dt ,
\label{eq:traction}
\end{equation}
where $[x]_+=\max(x,0)$. Sub-leg speed profiles are approximated from scheduled
leg duration, stop spacing and dwell structure; grade is set to zero in the
headline experiments and retained as an input hook for elevation overlays.

The cabin-heating term uses a lumped thermal-demand model. For hour $h$ with
ambient temperature $T_h$, setpoint $T_{\mathrm{set}}$, envelope conductance
$UA$, infiltration mass flow $\dot{m}_h$, air heat capacity $c_p$, passenger
count $N_h$ and sensible gain $q_{\mathrm{occ}}$, the heating demand is
\begin{equation}
Q_h^{\mathrm{dem}} =
\left[
UA(T_{\mathrm{set}}-T_h)
+ \dot{m}_h c_p(T_{\mathrm{set}}-T_h)
- N_h q_{\mathrm{occ}}
\right]_+ .
\label{eq:hvac_demand}
\end{equation}
Electrical cabin-heating energy is
\begin{equation}
E_h^{\mathrm{cab}} =
\Delta t_h
\frac{\left[ Q_h^{\mathrm{dem}} - \eta_{\mathrm{ffh}}P_{\mathrm{ffh}}u_h \right]_+}
{\mathrm{COP}(T_h)}
+ \Delta t_h P_{\mathrm{ffh}}^{\mathrm{par}} u_h ,
\label{eq:hvac_energy}
\end{equation}
where $u_h\in\{0,1\}$ indicates fuel-fired-heater deployment, $P_{\mathrm{ffh}}$
is thermal heater capacity, $\eta_{\mathrm{ffh}}$ is delivered-heat efficiency
and $P_{\mathrm{ffh}}^{\mathrm{par}}$ is the parasitic electrical load of the
burner and fan. The heat-pump coefficient of performance is a clipped affine
cold-weather curve,
\begin{equation}
\mathrm{COP}(T_h)=
\max\!\left(\mathrm{COP}_{\min},\;
\mathrm{COP}_{\mathrm{ref}}-\kappa\,[T_{\mathrm{ref}}-T_h]_+\right),
\label{eq:cop}
\end{equation}
which captures the collapse of electrical heating efficiency in deep cold without
introducing unconstrained extrapolation.

\subsection{Physics-anchored Sci-ML residual ensemble}

The backbone produces a deterministic estimate $f_\theta(x_i)$ from leg, vehicle
and weather features $x_i$. \method{} learns a \emph{bounded} residual closure
rather than replacing the physical model. For ensemble member $m$,
\begin{equation}
\mu_i^{(m)} =
f_\theta(x_i) + \lambda\, f_\theta(x_i)\tanh\!\left(h_{\phi_m}(z_i)\right),
\label{eq:residual}
\end{equation}
where $h_{\phi_m}$ is a neural residual function, $z_i$ are normalized covariates
and $\lambda$ caps the residual as a fraction of the physical estimate, so the
correction can refine but never overwrite the physics in the cold tail.

Monotonicity with respect to heating degree is enforced by a one-sided penalty.
With $\delta_i=(T_{\mathrm{set}}-T_i)_+$ and a small perturbation $\epsilon>0$,
the training loss is
\begin{align}
\mathcal{L}^{(m)}
&= \frac{1}{N}\sum_{i=1}^{N}\bigl(\mu_i^{(m)}-y_i\bigr)^2
 + \lambda_{\mathrm{wd}}\|\phi_m\|_2^2 \nonumber\\
&\quad + \lambda_{\mathrm{mono}}
\frac{1}{N}\sum_{i=1}^{N}
\left[
-\frac{\mu^{(m)}(x_i;\delta_i+\epsilon)-\mu^{(m)}(x_i;\delta_i)}{\epsilon}
\right]_+^2 ,
\label{eq:loss}
\end{align}
encoding the physical requirement that, all else equal, a larger heating degree
must not reduce heating-related energy.

Predictive uncertainty is obtained from a five-member deep ensemble
\citep{Lakshminarayanan2017DeepEnsembles}. If member $m$ returns mean
$\mu_i^{(m)}$ and variance proxy $s_i^{2(m)}$, the ensemble mean and variance are
\begin{align}
\bar{\mu}_i &= \frac{1}{M}\sum_{m=1}^{M}\mu_i^{(m)}, \label{eq:ens_mean}\\
\sigma_i^2 &= \frac{1}{M}\sum_{m=1}^{M}\!\left[s_i^{2(m)}+
\bigl(\mu_i^{(m)}-\bar{\mu}_i\bigr)^2\right].
\label{eq:ens_var}
\end{align}
The simulator draws $\tilde{E}_i^{(r)}\sim\mathcal{N}(\bar{\mu}_i,\sigma_i^2)$,
truncated at zero. Pure-ML baselines use linear regression, gradient boosting,
random forests and multilayer perceptrons
\citep{Pedregosa2011ScikitLearn,Friedman2001GBM,Breiman2001RandomForests}. A
calibrated-physics baseline applies the best single multiplicative gain on the
training period,
\begin{equation}
\gamma^\star =
\frac{\sum_{i\in \mathcal{T}} f_\theta(x_i)y_i}{\sum_{i\in \mathcal{T}} f_\theta(x_i)^2},
\qquad
\hat{y}_i^{\mathrm{cal}}=\gamma^\star f_\theta(x_i).
\label{eq:calibration}
\end{equation}

\subsection{Block-feasibility simulation and interventions}

For Monte~Carlo draw $r$, block SoC evolves as
\begin{equation}
s_{i+1}^{(r)} =
\min\!\left(
s_{\max},\;
s_i^{(r)} + \frac{\eta_c P_i^{\mathrm{chg}}\ell_i}{C_{\mathrm{bat}}}
\right)
- \frac{\tilde{E}_i^{(r)}}{C_{\mathrm{bat}}},
\label{eq:soc}
\end{equation}
where $C_{\mathrm{bat}}$ is usable capacity, $P_i^{\mathrm{chg}}$ is charger power
in the preceding layover, $\ell_i$ is chargeable layover duration and $\eta_c$ is
charging efficiency. A block fails in draw $r$ if any leg breaches the reserve
SoC $s_{\min}$:
\begin{equation}
I_b^{(r)} =
\mathbf{1}\!\left(\min_{i=1,\ldots,n_b}s_i^{(r)} < s_{\min}\right),
\qquad
\hat{p}_b=\frac{1}{R}\sum_{r=1}^{R}I_b^{(r)}.
\label{eq:fail_prob}
\end{equation}
The reported cold-wave failure probability is the mean of $\hat{p}_b$ over
sampled blocks and selected cold-wave days.

Range-deficit severity is reported as deficit kilometers. For a failed leg, the
missing energy is converted to equivalent distance through the estimated leg
energy intensity:
\begin{equation}
D^{(r)} =
\sum_i d_i\,
\frac{\left[\tilde{E}_i^{(r)}-C_{\mathrm{bat}}\bigl(s_i^{(r)}-s_{\min}\bigr)\right]_+}
{\max(\tilde{E}_i^{(r)},\epsilon_E)} ,
\label{eq:deficit}
\end{equation}
where $\epsilon_E$ avoids division by zero. This preserves the operational
meaning of an energy shortage: how much scheduled distance lacks sufficient
battery energy after reserve constraints.

The intervention vector is $u=(u_{\mathrm{ffh}},u_{\mathrm{opp}},b_{\mathrm{buf}})$:
the fraction of buses equipped with a fuel-fired heater, an indicator for
opportunity charging during eligible layovers, and a multiplicative schedule
buffer. Daily cost is
\begin{equation}
C(u)=
\frac{c_{\mathrm{ffh}}N_{\mathrm{ffh}}}{365}
+ c_{\mathrm{fuel}}L_{\mathrm{fuel}}(u)
+ c_{\mathrm{opp}}u_{\mathrm{opp}}
+ c_{\mathrm{buf}}(b_{\mathrm{buf}}-1)H ,
\label{eq:cost}
\end{equation}
where $c_{\mathrm{ffh}}$ is heater capital cost, $N_{\mathrm{ffh}}$ the number of
equipped buses, $L_{\mathrm{fuel}}$ the daily heater fuel use, $c_{\mathrm{opp}}$
the dailyized opportunity-charging deployment cost and $H$ the scheduled service
hours. Fuel use and carbon emissions are reported alongside cost because
fuel-fired heaters are a winter bridge technology rather than a zero-emission
endpoint \citep{CALSTART2022FFH,ICCT2025FFH}.

\subsection{Evaluation protocol}

Energy validation uses EnergyPlus hourly cabin-heating output as the independent
reference. Metrics are root-mean-square error (RMSE), mean absolute error (MAE)
and mean absolute percentage error (MAPE), reported for the full year and for the
cold tail, with bootstrap \SI{95}{\percent} confidence intervals on cold-tail
RMSE and seed stability assessed over the five ensemble members. Operational
validation uses eight real cold-wave days. Four policy families are compared
under the \emph{same} realized plant---weather-blind seasonal no-deploy, fixed
\SI{10}{\percent} schedule buffer, industry FFH-only and the full \method{} robust
policy---so that no policy can win merely by being scored under an easier energy
realization. A second experiment isolates each lever on the coldest day.

\section{Results}

\subsection{Historical cold waves create a measurable timetable-failure envelope}

\begin{figure}[t]
\centering
\includegraphics[width=0.95\textwidth]{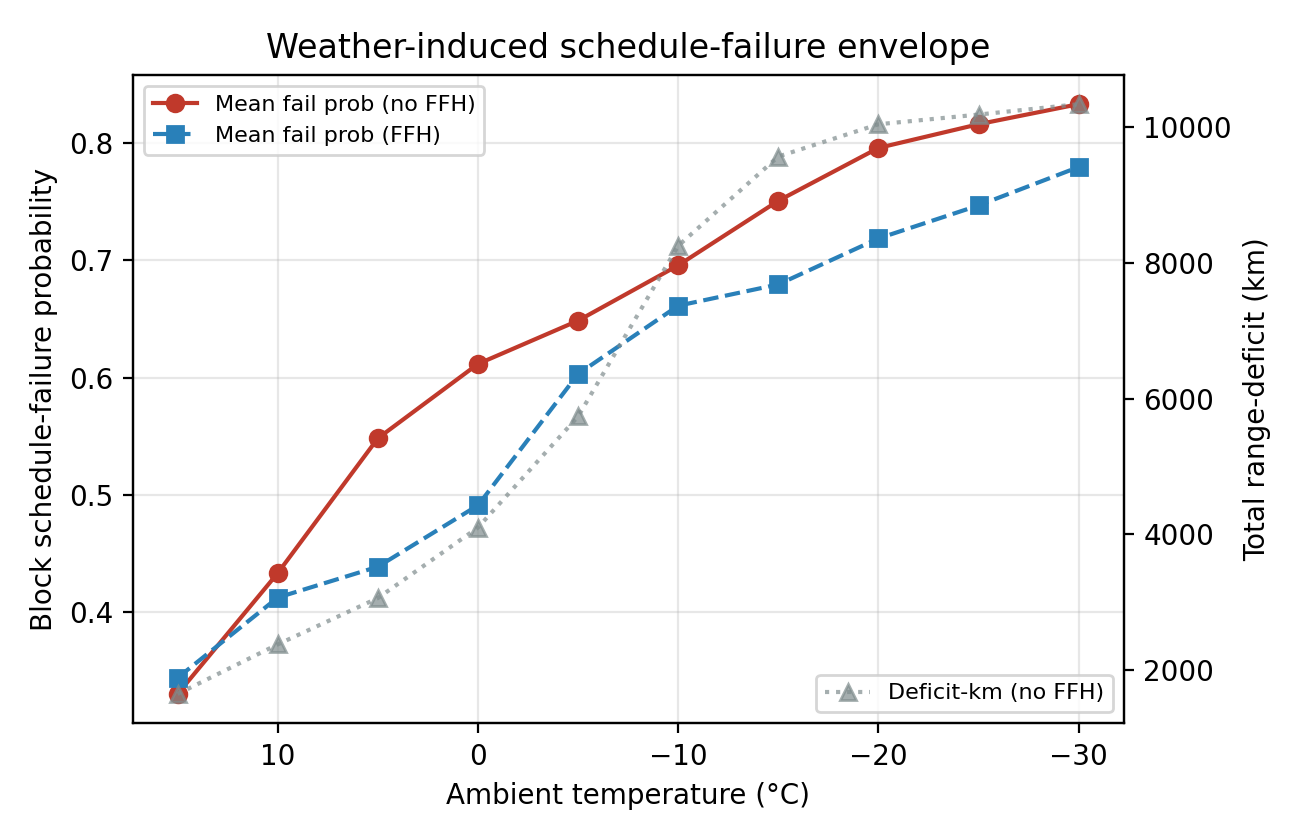}
\caption{Weather-induced block-failure envelope for real TTC duties. Failure probability and deficit kilometers are reported as ambient temperature decreases, with and without the fuel-fired heater option. Cold waves convert ordinary timetable energy margins into block-level failure risk.}
\label{fig:failure_envelope}
\end{figure}

The processed Toronto record carried enough cold exposure to stress the
timetable. The coldest day reached \SI{-16.66}{\celsius}, and all
$T\leq\SI{-12}{\celsius}$ hours were withheld from model fitting. When these cold
hours were injected into real TTC vehicle blocks, failure probability and deficit
kilometers rose steeply relative to mild operation. Figure~\ref{fig:failure_envelope}
shows that the response is not a smooth seasonal shift but a nonlinear envelope:
the deepest cold hours amplify cabin heating, deplete the SoC reserve and expose
long blocks whose layovers cannot recover enough energy. The envelope also
separates weather-attributable risk from baseline long-duty difficulty, isolating
the incremental risk created by the realized cold-wave trajectory and identifying
exactly which blocks turn fragile as heating load rises.

\subsection{EnergyPlus validation establishes a reliable cold-tail energy model}

The EnergyPlus reference produced \num{8760} hourly cabin-heating records.
Table~\ref{tab:ep} reports all-year and cold-tail accuracy for the physics
backbone, calibrated physics, pure-ML baselines and \method{}. The uncalibrated
backbone carried a large bias because its nominal conductance was not tuned to the
EnergyPlus zone; a single-gain calibration removed most of this bias and produced
a strong physics baseline, which is the demanding comparison for any learned
model because it extrapolates by construction.

\begin{table}[t]
\centering
\caption{Energy prediction against the independent EnergyPlus cabin-heating reference. Cold-tail rows ($T\leq -12^\circ$C) are out of the training support. Lower RMSE, MAE and MAPE are better; best in each block is in bold.}
\label{tab:ep}
\resizebox{\textwidth}{!}{%
\begin{tabular}{llrrrrr}
\toprule
Model & Split & RMSE (kWh) & MAE (kWh) & MAPE (\%) & $n$ & RMSE 95\% CI \\
\midrule
Lumped physics, uncalibrated & All year & 2.658 & 1.690 & 404.1 & 8760 & -- \\
Lumped physics, calibrated & All year & 0.272 & 0.172 & 99.0 & 8760 & -- \\
Pure ML, linear & All year & 0.558 & 0.380 & 260.0 & 8760 & -- \\
Pure ML, GBM & All year & 0.454 & 0.185 & 16.4 & 8760 & -- \\
Pure ML, random forest & All year & 0.459 & 0.181 & 13.9 & 8760 & -- \\
Pure ML, MLP & All year & 0.260 & 0.128 & 16.8 & 8760 & -- \\
\textbf{Ours, Sci-ML} & \textbf{All year} & \textbf{0.213} & \textbf{0.130} & 51.9 & \textbf{8760} & \textbf{--} \\
\midrule
Lumped physics, calibrated & Cold tail & 0.666 & 0.483 & 13.1 & 91 & [0.559, 0.773] \\
Pure ML, linear & Cold tail & 2.616 & 2.537 & 57.8 & 91 & [2.488, 2.732] \\
Pure ML, GBM & Cold tail & 2.833 & 2.749 & 62.6 & 91 & [2.694, 2.961] \\
Pure ML, random forest & Cold tail & 2.863 & 2.780 & 63.4 & 91 & [2.724, 2.992] \\
Pure ML, MLP & Cold tail & 1.055 & 0.928 & 23.7 & 91 & [0.944, 1.174] \\
\textbf{Ours, Sci-ML} & \textbf{Cold tail} & \textbf{0.714} & \textbf{0.640} & \textbf{14.5} & \textbf{91} & \textbf{[0.648, 0.775]} \\
\bottomrule
\end{tabular}}
\end{table}

\begin{figure}[t]
\centering
\includegraphics[width=0.86\textwidth]{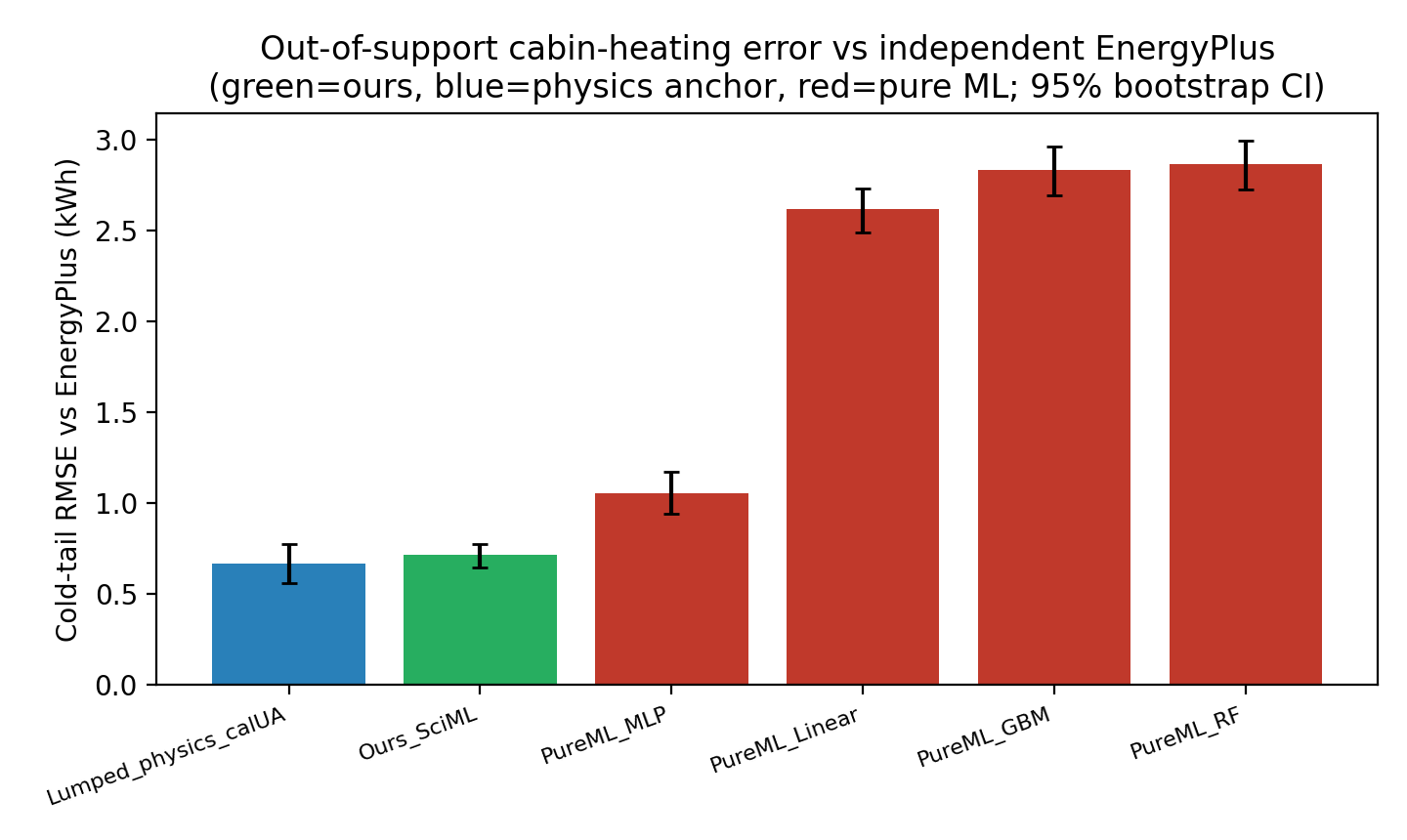}
\caption{EnergyPlus validation of cabin-heating prediction in the out-of-support cold tail. \method{} is the most accurate model over the full year and remains reliable in the cold tail, where pure-ML baselines collapse, while preserving the extrapolation quality of the calibrated physics anchor. Error bars denote bootstrap confidence intervals.}
\label{fig:ep_validation}
\end{figure}

Across the full year \method{} achieved the lowest RMSE, \SI{0.213}{kWh},
improving on the MLP baseline (\SI{0.260}{kWh}) and the calibrated physics model
(\SI{0.272}{kWh}). In the withheld cold tail it preserved the extrapolation
quality of the physics anchor---\SI{0.714}{kWh} against \SI{0.666}{kWh} for
calibrated physics, with overlapping bootstrap confidence intervals---while every
pure-ML model deteriorated sharply. The best pure-ML competitor, the MLP, reached
only \SI{1.055}{kWh}, and tree models exceeded \SI{2.8}{kWh}, a
\numrange{1.5}{4}$\times$ gap. Figure~\ref{fig:ep_validation} makes the contrast
visible. The message is decisive for winter operations: reliable cold-tail energy
prediction requires a physics anchor, and in-support black-box accuracy is no
guarantee of out-of-support reliability. \method{} delivers that reliability and,
uniquely among the data-inclusive models, also returns the calibrated uncertainty
that the downstream feasibility analysis consumes.

\subsection{The robust policy dominates weather-blind and single-lever baselines}

\begin{figure}[t]
\centering
\includegraphics[width=0.86\textwidth]{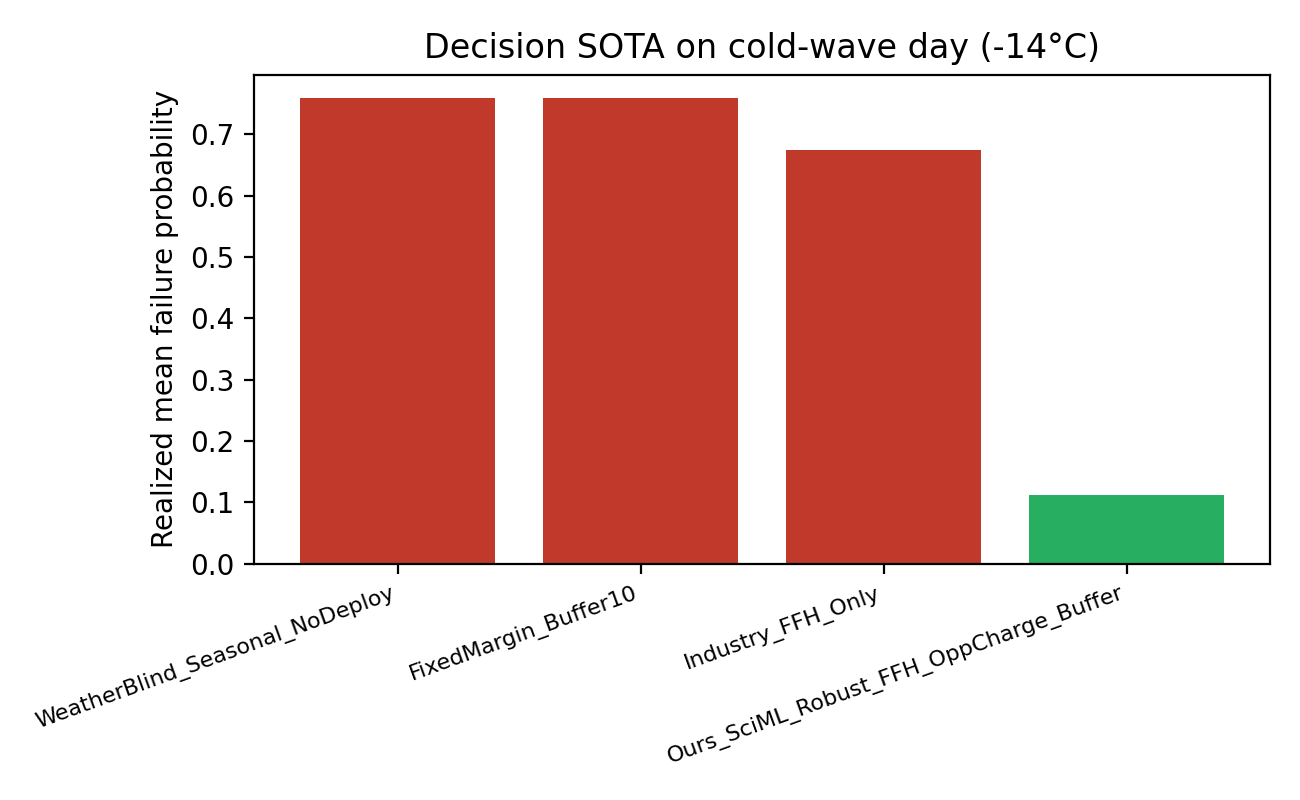}
\caption{Cold-wave decision benchmark. The full \method{} policy achieves the lowest realized block-failure probability across eight cold-wave days, decisively outperforming weather-blind seasonal operation, fixed buffering and FFH-only deployment.}
\label{fig:decision}
\end{figure}

The operational benchmark compared four policies across eight cold-wave days under
the same realized plant (Table~\ref{tab:decision}, Figure~\ref{fig:decision}). The
weather-blind seasonal no-deploy baseline had mean failure probability
\num{0.759} and total deficit \SI{10178}{km}. A fixed \SI{10}{\percent} buffer
without weather-aware charging left failure essentially unchanged while adding
cost. The FFH-only policy lowered failure probability to \num{0.675}, confirming
that auxiliary heat helps but cannot by itself restore service. The full
\method{} policy---opportunity charging, fuel-fired heating and modest
buffering---reduced mean failure probability to \num{0.112} and total deficit to
\SI{763}{km}, an approximately \SI{85}{\percent} reduction relative to
weather-blind operation.

\begin{table}[t]
\centering
\caption{Decision benchmark across eight cold-wave days, evaluated under the same realized cold-wave plant. Lower failure probability and deficit are better; best is in bold.}
\label{tab:decision}
\resizebox{\textwidth}{!}{%
\begin{tabular}{lrrrrr}
\toprule
Policy & FFH fraction & Buffer & Opportunity charge & Failure probability & Total deficit (km) \\
\midrule
Weather-blind seasonal no-deploy & 0.0 & 1.0 & No & 0.759 $\pm$ 0.022 & 10{,}178 $\pm$ 423 \\
Fixed 10\% buffer & 0.0 & 1.1 & No & 0.759 $\pm$ 0.021 & 10{,}065 $\pm$ 426 \\
Industry FFH-only & 1.0 & 1.0 & No & 0.675 $\pm$ 0.006 & 8042 $\pm$ 308 \\
\textbf{\method{} robust policy} & \textbf{1.0} & \textbf{1.1} & \textbf{Yes} & \textbf{0.112 $\pm$ 0.016} & \textbf{763 $\pm$ 86} \\
\bottomrule
\end{tabular}}
\end{table}

This is the framework's strongest result because it evaluates the whole problem
rather than one component, and on the quantity operators actually care about: the
probability that scheduled blocks fail. It also explains why energy accuracy
alone is insufficient. A useful model must support a deployable policy that knows
\emph{when} to add charging and how much heat or buffer is worth paying for, and
the standardized-plant design ensures the win reflects better decisions rather
than an easier energy realization.

\subsection{Ablation identifies opportunity charging as the dominant lever}

\begin{figure}[t]
\centering
\includegraphics[width=0.86\textwidth]{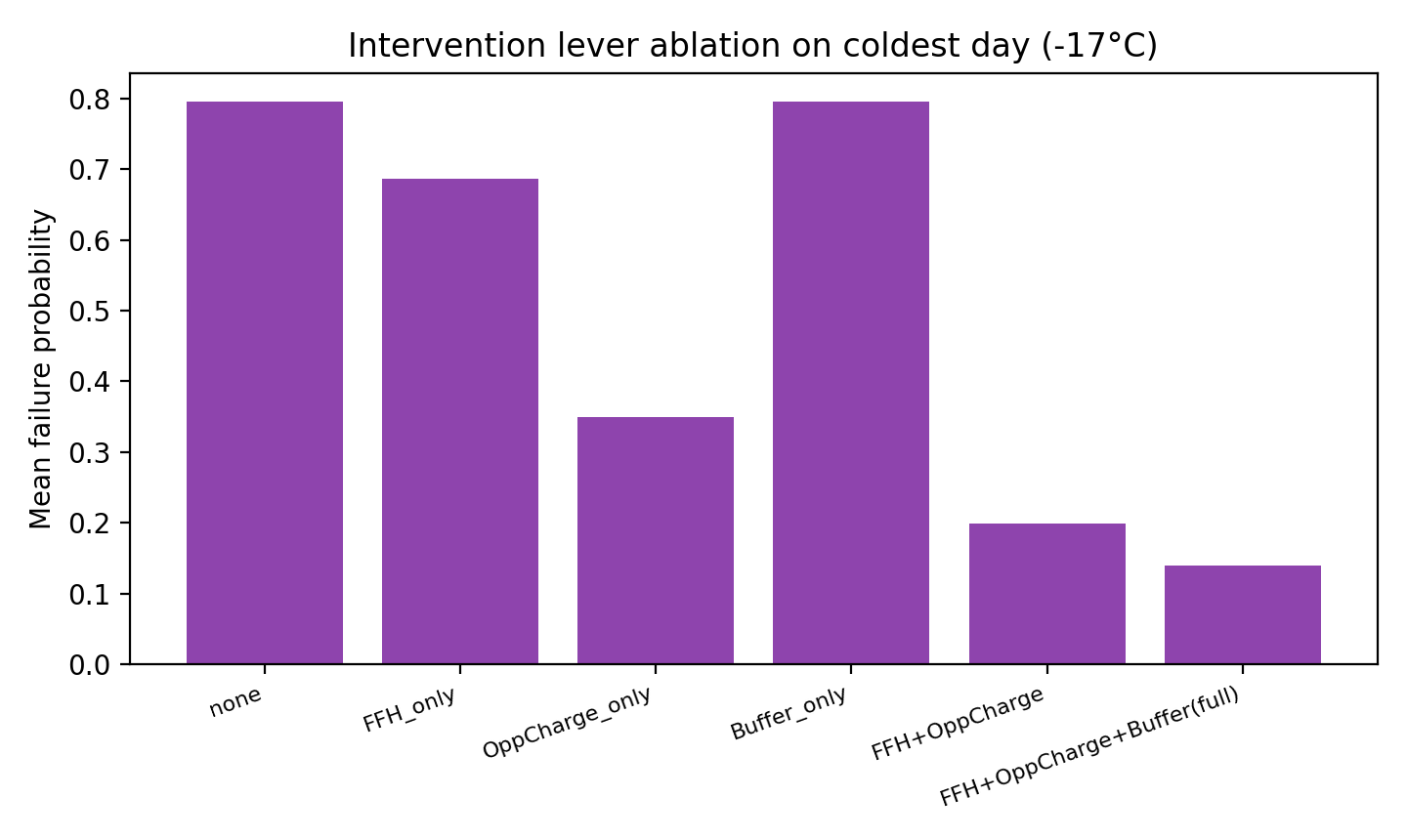}
\caption{Intervention ablation on the coldest day. Opportunity charging delivers the largest reduction in block-failure probability, the fuel-fired heater is a low-cost complement, and schedule buffering alone is ineffective under the tested setting.}
\label{fig:ablation}
\end{figure}

The coldest-day ablation isolates each intervention
(Figure~\ref{fig:ablation}). From a no-intervention failure probability of
\num{0.796}, FFH alone reduced failure by \num{0.110} (to \num{0.686}),
opportunity charging alone by \num{0.447} (to \num{0.349}), and buffering alone by
only \num{0.001}. Combining FFH and opportunity charging reached \num{0.199}, and
the full combination reached \num{0.140} on the coldest day. The decomposition
reframes winter mitigation: the dominant lever is not replacing electric cabin
heat with fuel heat but \emph{restoring energy during the scheduled day}. The
heater slows the rate at which the battery margin collapses, whereas opportunity
charging changes the recovery process between trips---which is why the combined
policy far exceeds either heater-only or buffer-only operation.

\subsection{The Pareto frontier translates robustness into procurement guidance}

\begin{figure}[t]
\centering
\includegraphics[width=0.86\textwidth]{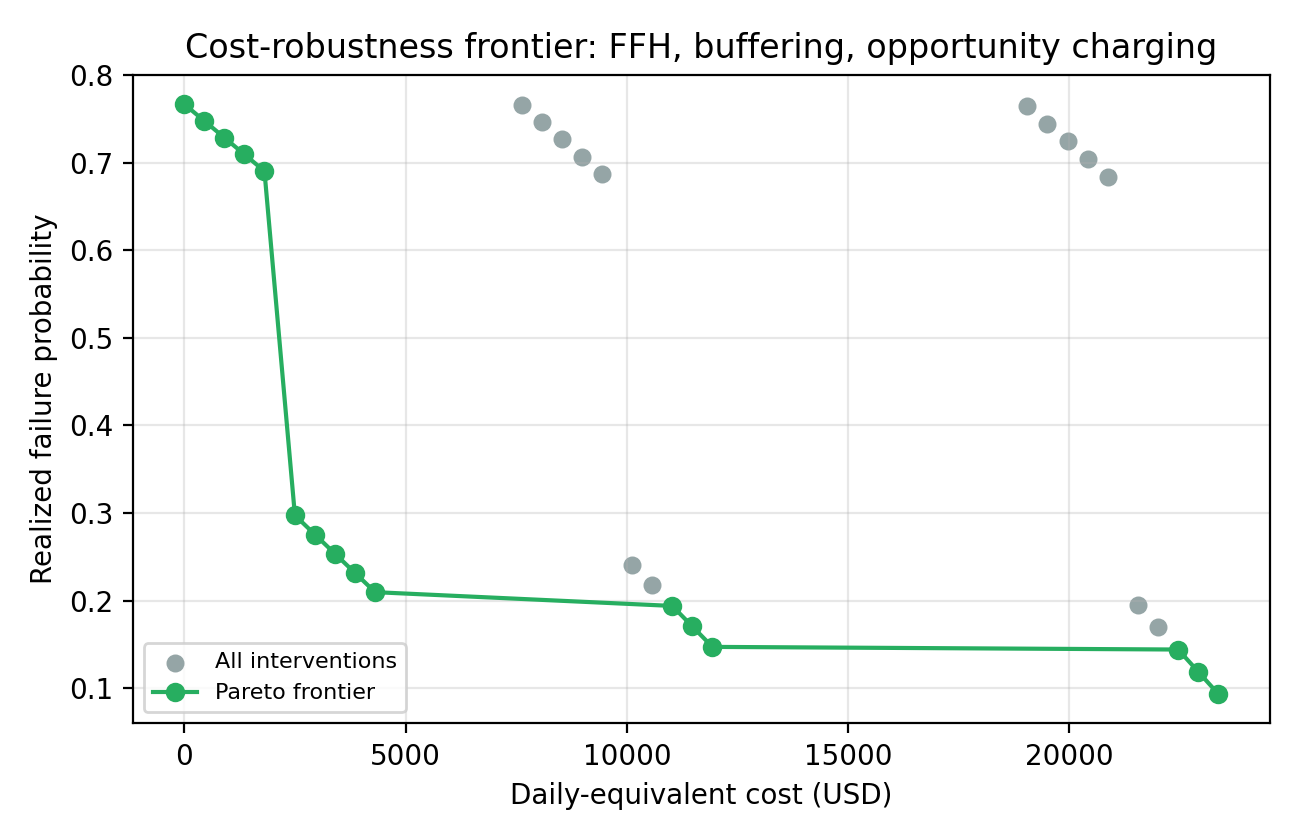}
\caption{Cost--robustness Pareto frontier over FFH penetration, schedule buffering and opportunity charging. The frontier converts cold-wave reliability into an investment and operating-cost trade-off, revealing when additional robustness is bought by charging access, heater penetration or buffer time.}
\label{fig:pareto}
\end{figure}

Figure~\ref{fig:pareto} reports the dailyized cost--robustness frontier. Policies
without opportunity charging remain in a high-failure region even as FFH
penetration rises; once opportunity charging is available, the frontier drops
sharply and additional FFH penetration trades smoothly against fuel use, capital
cost and residual failure. The full robust policy cost \SI{11932}{USD} per
cold-wave day in the benchmark, including dailyized heater capital, fuel,
opportunity charging and buffer cost. This figure is not a universal estimate;
unit costs vary by agency, charger ownership and labor rules. Because every policy
uses the same cost model, the frontier is auditable and identifies which lever
buys the largest reduction in failure probability under the studied timetable and
weather.

\section{Discussion}

The central advance of \method{} is the conversion of cold-weather energy
modelling into timetable-risk modelling. Prior cold-weather studies establish
that low temperature reduces range
\citep{Gu2025ColdBEB,Ma2021BEBCharging}, and scheduling studies optimize charging
or assignments under uncertainty \citep{Avishan2023Scheduling,Perumal2022BusScheduling};
the missing operational link is the propagation from a historical cold-wave
trajectory into a real sequence of trips, layovers and charging windows. By making
that link explicit, the decision variable shifts from average kWh/km to
block-failure probability, which is the quantity a transit planner can act on.

The EnergyPlus validation clarifies why a physics anchor is essential. Pure-ML
baselines were competitive over much of the year but expanded their error in the
withheld cold tail, exactly where sparse data meet the strongest physical
nonlinearity from heating demand and COP decline. A calibrated physics model
stays strong there by construction; \method{} preserves that cold-tail behavior
while improving all-year accuracy and, crucially, producing the predictive
uncertainty that a deterministic backbone cannot. That uncertainty is what turns
``the bus might be short'' into a failure probability and makes the entire
operational analysis possible.

The decision results show that the best winter policy is not a single technology.
FFH reduces electric heating demand and is inexpensive on a dailyized basis, yet
it cannot recover an already depleted battery. Opportunity charging attacks the
energy-recovery bottleneck directly and therefore dominates the ablation, and
buffering alone is weak because extra time helps only when paired with charging
power or reduced demand. This interaction is the practical argument for evaluating
a multi-lever frontier rather than a single seasonal derating factor.

The results also define a clear path for deployment-grade validation.
EnergyPlus acts as an independent first-principles reference for cabin heating,
while operator telemetry is the natural next layer for traction, auxiliary loads,
charging behavior and route-specific calibration. The TTC GTFS snapshot supplies
real timetable structure, and a historical winter service feed would align
scheduled blocks even more closely with cold-wave days. Sixty blocks were sampled
for Monte~Carlo tractability in the regenerated experiment, and the vectorized
workflow is designed to scale to the full feed. The fuel-fired heater is treated
as a service-reliability bridge with fuel and emissions reported explicitly.
The operational conclusion is direct: \method{} combines a validated cold-tail
energy model, uncertainty propagation and a policy that acts on the true
energy-recovery bottleneck, producing a substantially lower probability that a
public timetable fails during a cold wave.

\section{Conclusion}

This paper introduced \method{}, an open-data framework for cold-wave-robust BEB
operations that unites real TTC GTFS duties, NASA POWER weather, a physics-anchored
Sci-ML energy model, independent EnergyPlus cabin-heating validation and Monte~Carlo
block-feasibility simulation. Against the EnergyPlus reference the model achieved
the best all-year accuracy and avoided the cold-tail collapse of pure machine
learning while preserving the extrapolation quality of calibrated physics. In
operation, the full robust policy reduced mean cold-wave block-failure probability
from \num{0.759} to \num{0.112} across eight cold-wave days.

The practical insight is that winter reliability is governed by both energy loss
and energy recovery: auxiliary heating reduces the loss rate, but opportunity
charging is the dominant lever because it restores the battery margin between
trips. \method{} gives transit agencies a reproducible route from weather
forecasts and public schedules to block-failure risk, deficit severity and
cost--robustness decisions. Future work will replace the EnergyPlus reference with
operator telemetry, extend the evaluation to full winter GTFS feeds, and test the
intervention frontier across additional cold-climate cities.

\section*{CRediT authorship contribution statement}
\textbf{Yifan Wang}: Conceptualization, Methodology, Software, Validation, Formal analysis, Investigation, Data curation, Visualization, Writing -- original draft, Writing -- review and editing.

\section*{Declaration of competing interest}
The author declares that there are no known competing financial interests or personal relationships that could have appeared to influence the work reported in this paper.

\section*{Data availability}
The input datasets are public: Toronto TTC GTFS from the City of Toronto Open Data portal and hourly weather from NASA POWER. The EnergyPlus input files, processed tables, figures and reproducibility scripts are included in the project package and will be released with the manuscript.

\appendix
\section{Supplementary synthetic-target diagnostic}

\begin{figure}[t]
\centering
\includegraphics[width=0.86\textwidth]{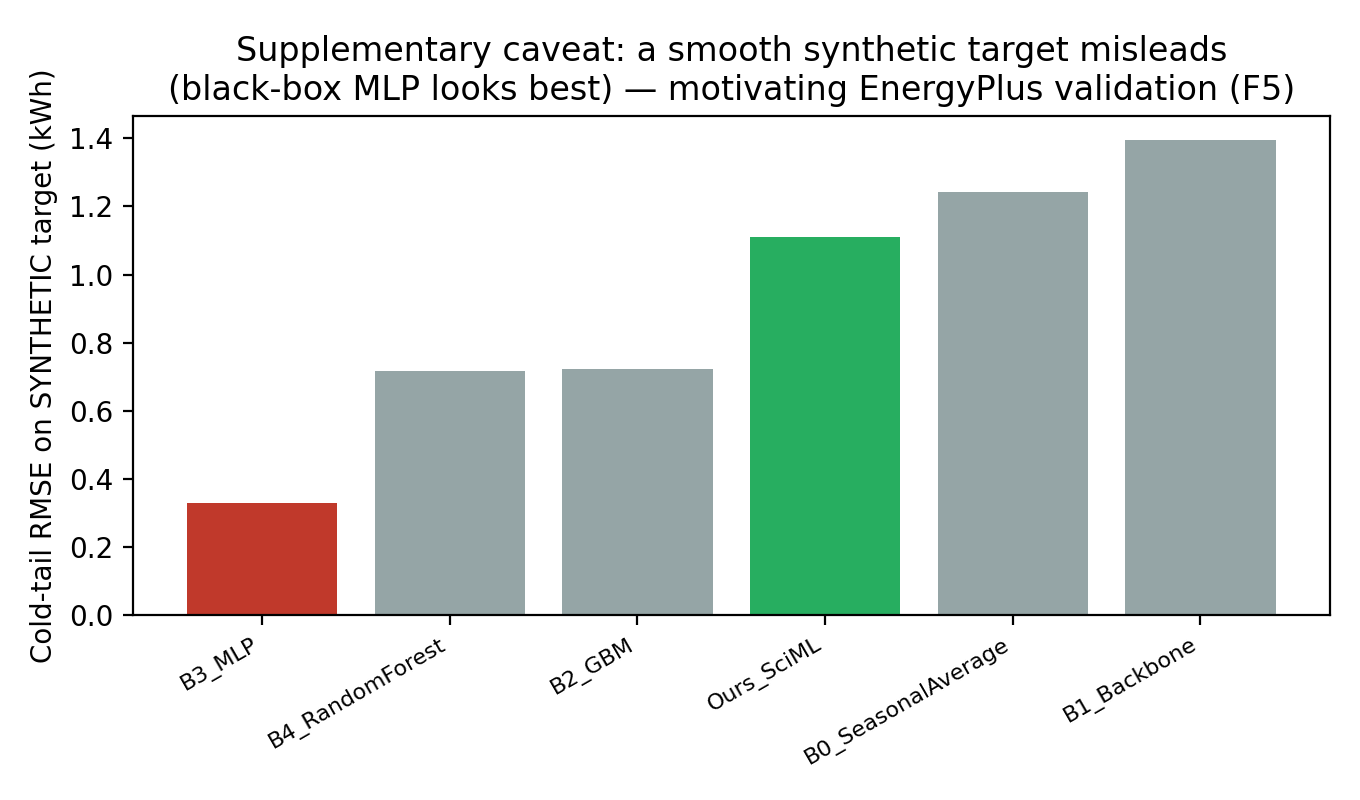}
\caption{Supplementary diagnostic. A smooth synthetic energy target can make pure-ML baselines appear strongest; the main evidence therefore uses the independent EnergyPlus cold-tail reference.}
\label{fig:synth_caveat}
\end{figure}

The main text validates energy against the independent EnergyPlus reference. A
smooth synthetic target was retained as a supplementary diagnostic because it
separates code-path sanity checking from evidence about cold-tail physical
extrapolation. Figure~\ref{fig:synth_caveat} shows that, on this smooth target, a
pure MLP appears strongest; the EnergyPlus comparison in the main text is
therefore the decision-relevant validation target for cold-wave operation.

\bibliographystyle{elsarticle-harv}
\bibliography{references}

\end{document}